\documentclass{ptptex}
\usepackage{amsmath,amsfonts,epsf,graphicx}
\newcommand{\tr}{\operatorname{tr}}
\newcommand{\ve}{\varepsilon}

\renewcommand{\Re}{\operatorname{Re}}

\begin{document}

\markboth{J.~P.~Keating {\it et al.}}{Periodic orbit bifurcations
and scattering time delay fluctuations}

\title{Periodic orbit bifurcations and scattering time delay fluctuations}

\author{J.~P.~Keating$^1$, A.~M.~Ozorio de Almeida$^2$,
S.~D.~Prado$^3$, M.~Sieber$^1$, and R.~Vallejos$^2$}

\inst{$^1$ School of Mathematics, University of Bristol BS8 1TW, UK \\
$^2$ Centro Brasileiro de Pesquisas F\'{\i}sicas, Rua
 Xavier Sigaud 150, 22290-180 Rio de Janeiro, RJ, Brazil \\
$^3$ Instituto de F\'{\i}sica, Universidade Federal do Rio Grande
do Sul, 91501-970 Porto Alegre, RS, Brazil}

\abst{We study fluctuations of the Wigner time delay for open
(scattering) systems which exhibit mixed dynamics in the classical
limit. It is shown that in the semiclassical limit the time delay
fluctuations have a distribution that differs markedly from those
which describe fully chaotic (or strongly disordered) systems:
their moments have a power law dependence on a semiclassical
parameter, with exponents that are rational fractions. These
exponents are obtained from bifurcating periodic orbits trapped in
the system. They are universal in situations where sufficiently
long orbits contribute. We illustrate the influence of
bifurcations on the time delay numerically using an open quantum
map.}

\maketitle

\section{Introduction}

In closed systems, for example closed billiards, quantum energy
levels and eigenfunctions are related to classical periodic orbits
in the semiclassical limit via trace formulae.  In the case of the
energy level spectrum, the trace formula was first developed by
Gutzwiller \cite{Gutzwiller:book}.  Combining trace formulae with
statistical information about the classical dynamics
(e.g.~ergodicity) underpins the semiclassical theory of quantum
fluctuation statistics in closed systems.  For example, this
approach forms the basis of attempts to understand universality
and the connection with random matrix theory in the spectral
statistics of fully chaotic systems, and likewise the connection
with Poisson statistics in regular systems.

In closed systems that have a mixed (i.e.~partly regular and partly
chaotic) classical limit, some quantum fluctuation statistics are
well described semiclassically by weighted averages over the regular
and chaotic components.  However others, for example moments of the
fluctuations of the spectral counting function around the Weyl mean,
or moments of the fluctuations of eigenfunctions about the quantum
ergodic limit, are dominated in the semiclassical limit by classical
periodic orbits that are close to bifurcation, where their
contribution to the trace formulae is enhanced by a power of
Planck's constant that depends on the nature of the bifurcation in
question \cite{Berry:1998, Berry_1:2000, Keating:2001, Gutierrez:2006}.
When long orbits contribute it is assumed that one can average over all
of the generic bifurcations. There is then a competition as to which
dominates the semiclassical moment asymptotics: essentially the more
complicated a bifurcation is the larger is its contribution but the
smaller is its range of influence.  In the cases studied so far
\cite {Berry_1:2000, Keating:2001} it happens that for any given
moment there is a bifurcation which dominates the competition.
Different moments are dominated by different bifurcations.  The
result is that each moment scales semiclassically as a power of
Planck's constant. These exponents are universal.  They take the
form of rational fractions whose values are given by simple
formulae. This differs markedly from the behaviour one sees in
either fully chaotic or fully regular systems.

Open (scattering) systems also exhibit universal quantum
fluctuation statistics, for example in the conductance and the
time delay, that depend on whether the classical dynamics is
regular or chaotic \cite{Smilansky:1991}. In chaotic scattering
systems the fluctuation statistics coincide with those of random
matrix theory.  The question then arises as to what happens in
open systems in which the classical dynamics is mixed: is there
any analogue of the bifurcation-dominated fluctuation statistics
found in closed systems?  One might initially think not, because
the classical trajectories that underlie the semiclassical
expression for the $S$-matrix are scattering orbits, not periodic
orbits. However, it turns out that some quantities related to
scattering can be re-expressed directly in terms of the periodic
orbits trapped inside the scattering region. For example, if the
scattering system is a billiard with holes cut in the perimeter,
these quantities can be expressed either in terms of orbits
entering and exiting through the holes, or in terms of periodic
orbits that never hit the holes.  Thus in mixed open systems
bifurcating periodic orbits can semiclassically dominate the
related fluctuation statistics in exactly the same way as in
closed systems, giving rise to new classes of universal scaling
exponents that cannot be described within a random matrix model.
One example where this was recently shown to be the case is the
conductance fluctuations in antidot lattices \cite{Keating:2005}.

Our purpose here is to point out that the Wigner time delay is
another general example.  Specifically, we argue that in mixed
open systems fluctuations in the time delay may be dominated by
classical periodic orbit bifurcations, and that when long orbits
contribute the fluctuation moments scale semiclassically with
universal exponents whose values, again rational fractions, are
related directly to those calculated previously. We illustrate our
theory with numerical computations for a class of quantum maps.

\section{The Wigner time delay}

The concept of time delay in quantum scattering was introduced by
Eisenbud \cite{Eisenbud:1948} and Wigner \cite{Wigner:1955} in the
context of one-channel spherical wave scattering.  Later, Smith
\cite{Smith:1960} extended the notion to the $M$-channel case by
introducing the lifetime matrix
\begin{equation}
Q_{ij}=-i\hbar\sum_{c=1}^{M}S_{ic}^\dagger\frac{d}{dE}S_{cj}(E),
\label{lifetime_matrix}
\end{equation}
where $S$ is the standard scattering matrix and the sum runs over
all $M$ open channels denoted by $c$.  The time delay is defined
to be the average of the eigenvalues of $Q$:
\begin{align} \label{wignertimedelay} \tau(E) & =- \frac{i
\hbar}{M} \tr S^\dagger \frac{d}{d E} S = -\frac{i \hbar}{M}
\frac{d}{d E} \log \det S.
\end{align}
It can be interpreted as the typical time spent by a scattered
particle in the interaction region.

The time delay turns out to be very closely related to the density
of states in a closed system. This allows one to use the well
developed semiclassical apparatus for the density of states of
closed systems to unravel features of open ones. The connection was
identified (independently \cite{Carvalho:2002}) by Friedel
\cite{Friedel:1952} and Lifshitz \cite{Lifshitz:1952}: the average
time delay (\ref{wignertimedelay}) is related to the difference
between the level density of an interacting Hamiltonian $H$ with
respect to a free or reference Hamiltonian $H_{0}$.

Friedel's formalism was used by Balian and Bloch
\cite{Balian:1974} to derive a semiclassical expression for the
time delay which is closely related to the Gutzwiller trace
formula for the density of states.  This splits into a smooth part
$\overline{\tau}(E)$ and a fluctuating part $\tau^{\text{fl}}(E)$
determined by periodic orbits:
\begin{equation} \tau(E) = \frac{2\pi\hbar}{M} \rho(E) \approx
\frac{2\pi\hbar}{M} \left( \overline{\rho}(E)+ \rho^{\text{fl}}(E)
\right),  \label{cavities}
\end{equation}
where $\rho$ is the renormalized density of states. It has a
natural interpretation in the context of inside-outside duality
\cite{Doron:1992, Vallejos:1998}.

The smooth term on the right-hand side of (\ref{cavities}) can be
interpreted as the mean density of scattering resonances. It
represents the mean time spent in the scattering system. The
fluctuating term is given by
\begin{equation}
\rho^{\text{fl}}(E)  \approx  \frac{1}{\pi \hbar} \Re
\sum_{\gamma} \sum_{m=1}^\infty
\frac{T_\gamma}{\sqrt{|\det(M^m_\gamma - I)|}} \exp \left( i
\frac{m {\cal S}_\gamma}{\hbar} - i \frac{\pi}{2} m \mu_\gamma
\right) \label{td_fluct}\end{equation} where the sum runs over the
periodic orbits $\gamma$ which are trapped in the repeller and
their repetitions. $T_\gamma$, ${\cal{S}}_\gamma$, $M_\gamma$, and
$\mu_\gamma$ are respectively period, action, stability matrix and
Maslov index of the orbit $\gamma$.

The semiclassical analysis of the time delay is often based on the
relation to the trapped periodic orbits.   For example, in fully
chaotic open systems, statistical properties of these orbits can be
invoked to justify the use of random matrix theory to model the
fluctuations \cite{Eckhardt:1993,Vallejos:1998,Kuipers:2006}. It is
worth pointing out that all the emphasis so far has been placed on
systems in which the classical limit is fully chaotic. This is not
however the typical situation in actual experiments, where real
cavities or soft potentials lead to mixed dynamics. Our aim here is
to go beyond this idealization by incorporating the contributions
from orbit bifurcations.  This is important, because the
contributions are, in certain regimes, actually the dominant ones.

\section{The time delay for quantum maps}

We illustrate the general theory of the last section using quantum
maps, since in this case the relation between the time delay and
the trapped periodic orbits can be derived straightforwardly.

We start with a map that acts on a phase space corresponding to the
unit torus. Its quantization is defined by a unitary time evolution
operator $U$ with finite dimension $N$. The energy dependence of
Hamiltonian systems is simulated by including a phase factor
$\tilde{U} = e^{i \ve} U$ that depends on the quasi-energy $\ve$.
The map is then opened up by removing vertical strips from phase
space. These play the role of holes in the boundary of open
billiards.  The corresponding scattering matrix $S$ is an
$M$-dimensional matrix, where $M<N$ is the total number of position
states in the opening. It is the unitary matrix for the transition
from the opening onto itself after an arbitrary number of iterations
of the internal map \cite{Ozorio:2000}
\begin{equation} \label{smatrixmaps}
S(\ve) = \tilde{U}_{OO} + \sum_{n=0}^\infty \tilde{U}_{OI}
\tilde{U}_{II}^n \tilde{U}_{IO} \, .
\end{equation}
The quantities $\tilde{U}_{kl}$ with $k,l \in \{O,I\}$ are
restrictions of the evolution operator to the inside and/or
outside. They are defined by $\tilde{U}_{kl} = P_k \tilde{U}
P^T_l$, where $P_O$ and $P_I$ are the projection matrices onto the
opening and the interior, respectively. $P_O$ and $P_I$ have
dimensions $M \times N$ and $(N-M) \times N$, respectively, and
$P_O^T$ and $P_I^T$ are the corresponding transpose matrices. The
projection operators satisfy $P_O^T P_O + P_I^T P_I = I_N$, $P_O
P_O^T = I_M$, and $P_I P_I^T = I_{N-M}$ where $I_L$ denotes the $L
\times L$ unitary matrix.

The sum over $n$ in (\ref{smatrixmaps}) can be performed and the
$S$-matrix written as
\begin{equation}
S(\ve) = P_O \frac{1}{I_N - \tilde{U} P_I^T P_I} \tilde{U} P_O^T \; .
\end{equation}
Planck's constant for quantum torus maps is given by $\hbar = (2
\pi N)^{-1}$, and so the time delay has the form
\begin{equation} \label{tau}
\tau(\ve)   = -\frac{i}{2\pi N M} \frac{d}{d \ve} \log \det S.
\end{equation}

In order to derive a formula for the time delay in terms of the
periodic orbits of the map, it is useful to apply an identity for
determinants due to Jacobi. Let $A$ be an $N \times N$ matrix given
in terms of the auxiliary block matrices $B, C, D$ and $E$, and
$A^{-1}$ its inverse matrix given in terms of $W,X,Y$ and $Z$:
\begin{equation}
A      = \begin{pmatrix} B & C \\ D & E \end{pmatrix} \quad
A^{-1} = \begin{pmatrix} W & X \\ Y & Z \end{pmatrix} \; .
\end{equation}
$B$ and $W$ are assumed to be square matrices with the same
dimension. Jacobi's determinant identity then states that $\det
B=\det Z \det A$.

Let us identify $A$ with the $N \times N$ matrix $[I_N - \tilde{U}
P_I^T P_I]^{-1} \tilde{U}$ and the subblock $B$ with $P_O A P_O^T =
S$. Then $Z$ follows as $P_I A^{-1} P_I^T$ and the Jacobi identity
leads to
\begin{equation} \label{detS}
\det S = \det( [I_N - \tilde{U} P_I^T P_I]^{-1} \tilde{U})
\det(P_I \tilde{U}^{-1} [I_N - \tilde{U} P_I^T P_I] P_I^T).
\end{equation}
A little elementary linear algebra then gives
\begin{equation}
\det S = \det \tilde{U} \; \frac{
\overline{\det(\tilde{U}_{II} - I_{N-M})}}{
\det(I_{N-M} - \tilde{U}_{II})} \; .
\end{equation}
For the evaluation of the derivative of the logarithm of $\det S$
we need
\begin{equation}
\frac{d}{d \ve} \log \det \tilde{U}
= \frac{d}{d \ve} \log e^{i N \ve} \det U = i N \; ,
\end{equation}
and
\begin{align}
\frac{d}{d \ve} \log \det (I_{N-M} - \tilde{U}_{II})
= - i \sum_{n=1}^\infty e^{i n \ve} \tr U_{II}^n \; .
\end{align}
This leads to the final result
\begin{equation}
\tau(\ve) =\frac{1}{M N} \left( \frac{N}{2 \pi} + \frac{1}{\pi}
\Re \sum_{n=1}^\infty e^{i n \ve} \tr U_{II}^n \right ),
\label{tracetau}
\end{equation}
which is in agreement with the general formula (\ref{cavities}). The
first term on the r.h.s. of (\ref{tracetau}) represents the mean
time spent in the scattering region, $\langle \tau(\ve) \rangle$,
while the second term is the fluctuating part of the time delay
$\tau^{\text{fl}}(\ve)$. It should be noted that
$\tau^{\text{fl}}(\ve)$ is written in terms of $\tilde{U}_{II}$
rather than $\tilde{U}$, the evolution operator for the unopened
map. The powers of traces of $U_{II}$ are semiclassically related to
the periodic orbits which lie completely in the interior of the open
map. If the traces are evaluated in the semiclassical approximation
one reproduces (\ref{cavities}) and (\ref{td_fluct}).

\section{Contribution of bifurcations to the time delay}

Generic systems have a phase space in which regular islands and
regions of chaotic sea coexist. One of the main characteristics of
mixed systems is the bifurcation of periodic orbits. Bifurcations
are events where different periodic orbits coalesce when
parameters of the system are varied. They are important in
semiclassical approximations because bifurcating orbits carry a
semiclassical weight that is higher than that of the isolated
(unstable) periodic orbits and sometimes even that of tori of
regular orbits.

Consider the Gutzwiller contribution of a periodic orbit $\gamma$
and its repetitions $m$. The amplitude in (\ref{td_fluct}) diverges
if $M^m_\gamma$ has an eigenvalue one, which happens at a
bifurcation. This is because periodic orbits are assumed to be
isolated in the derivation of the trace formula
\cite{Gutzwiller:book}. More specifically, the trace formula can be
derived by integrating over Poincar\'e sections perpendicular to
periodic orbits, and periodic orbits appear as stationary points of
these integrals. Consequently, the usual stationary phase
approximation breaks down when stationary point coalesce. The remedy
is to perform a uniform asymptotic expansion valid throughout the
bifurcation process \cite{Ozorio:1987, Sieber:1996, Schomerus:1997,
Schomerus:1998, Ozorio:book}. This is obtained by rederiving the
trace formula using the appropriate generating function for the
Poincar\'e map $\Phi(Q^{'},P)$ from ($Q,P$) to ($Q^{'}, P^{'}$) in
normal form coordinates.  For two-dimensional systems the
semiclassical contribution of a bifurcation to the density of states
is then given by
\begin{equation}
 \rho^{\text{fl}} \propto \frac{1}{\hbar^2}
\int d Q d P \exp{(i \Phi(Q,P)/\hbar)}. \label{oscillatory}
\end{equation}
As an example we take a saddle node bifurcation, $\Phi(Q,P) = P^2 +
x_1 Q + Q^3$, where $x_1$ depends on the energy or other system
parameters and vanishes at the bifurcation. The stationary points
occur for negative $x_1$ at $(Q,P) = (\pm \sqrt{-x_1},0)$. The
contribution (\ref{oscillatory})
at the bifurcation is $\propto \hbar^{-\beta}$, where
$\beta=7/6$. Away from the bifurcation (i.e.~when $-x_1/\hbar$ is
large) one obtains contributions of isolated orbits that are
$\propto \hbar^{-1}$. The two regimes are interpolated by an Airy
function.

Besides $\beta$ there are further exponents that are important for
the semiclassical influence of the bifurcation. They describe the
size of the parameter intervals over which the bifurcation is
semiclassically stronger than isolated periodic orbits. Consider
$\Phi(Q,P) = P^2 + x_1 Q + Q^3$ in the example. We can make the
exponent in the integral (\ref{oscillatory}) $\hbar$-independent
by scaling $Q = \tilde{Q} \hbar^{1/3}$, $P = \tilde{P}
\hbar^{1/2}$ and $x_1 = \tilde{x}_1 \hbar^{2/3}$.  Hence the
relevant $x_1$ interval scales like $\hbar^{\sigma_1}$, where
$\sigma_1=2/3$ in this example. The same analysis can be applied
to more complicated bifurcations which have more parameters $x_i$,
$i=1,\ldots,K$, in their normal form.  ($K$ denotes the
codimension of the bifurcation.)  One then obtains a
characteristic exponent $\sigma_i$ for every parameter $x_i$.
Because of this finite extension in parameter space, bifurcations
of higher codimension have to be taken into account even if only
one parameter is varied.

Generic bifurcations of periodic orbits are characterized by the
codimension $K$ and the repetition number $m$ of the orbit for
which the bifurcation occurs. A systematic investigation of the
influence of the different bifurcations on moments of the density
of states was carried out by Berry, Keating and Schomerus
\cite{Berry_1:2000}. This has subsequently been extended to
determine their influence on the statistics of wavefunctions
\cite{Keating:2001, Backer} and, more recently, on the moments of
the conductance fluctuations in antidot lattices
\cite{Keating:2005}.

Fluctuations of the Wigner time delay  can be characterized by
their moments, defined as
\begin{equation}
{\cal{M}}_{2k} = \left ( \frac{M}{2\pi\hbar}\right )^{2k}
\left\langle (\tau^{\text{fl}})^{2 k} \right\rangle_{E,X}
\label{moments}
\end{equation}
where $\langle \cdots \rangle_{E,X}$ denotes averaging over energy
and over parameter space. In any parameter interval of a system with
mixed dynamics infinitely many bifurcations occur, most of them for
very long periodic orbits. If $\hbar$ is small enough, then these
bifurcations are important.  To determine the influence of a
particular bifurcation with codimension $K$ and repetition number
$m$ on the moments of the time delay, we replace the average in
(\ref{moments}) by an average over parameters in the normal form.
Performing the scaling procedure \cite{Berry_1:2000} we can extract
the $\hbar$ dependence for the different bifurcations
${\cal{M}}_{2k,m,K} \sim \hbar^{-\eta_{k,m,K}}$ where $\eta_{k,m,K}
= 2 k \beta_{m,K} - \sum_{i=1}^K \sigma_{i,m,K}$. Hence we see that
the importance of a bifurcation depends on the quantity
$\eta_{k,m,K}$, which in turn depends on the characteristic
exponents $\beta_{m,K}$ and $\sigma_{i,m,K}$. The bifurcation that
is most important for a particular moment is that for which
$\eta_{k,m,K}$ is largest, and hence one finds that ${\cal{M}}_{2k}
\sim \hbar^{-\eta_k}$ where $\eta_k = \max_{m,K}( \eta_{k,m,K})$.
There is a different winner of this competition between bifurcations
for every $k$. It follows from (\ref{cavities}) that all the
different exponents for the time delay (\ref{moments}) coincide with
those for the corresponding moments of the density of states
\cite{Berry_1:2000, Keating:2005}.

\section{Numerical results}

We now illustrate some of the general ideas described above using
a family of perturbed cat maps \cite{Basilio, Boasman:1995}. These
are maps of the form
\begin{equation}
\begin{pmatrix} q' \\ p' \end{pmatrix}  =
\begin{pmatrix} 2 & 1 \\ 3 & 2 \end{pmatrix}
\begin{pmatrix} q \\ p \end{pmatrix}
+ \frac{\kappa}{2\pi} \cos(2\pi q)
\begin{pmatrix} 1 \\ 2 \end{pmatrix} {\mathrm{mod}} \; 1
\label{map}
\end{equation}
where $q$ and $p$ are coordinates on the unit two-torus, and are
taken to be a position and its conjugate momentum.  We will
concentrate on one particular bifurcation and investigate its
influence on the second moment of the time delay. Using
(\ref{moments}), (\ref{cavities}) and (\ref{tracetau}) one can write
\begin{align}
{\label{momentsmap}} {\cal{M}}_{2}\equiv & \frac{1}{2\pi}
\int^{2\pi}_{0}({\rho^{fl}}_{\Delta}(\ve))^{2} d\ve \approx
\frac{1}{2\pi^2} \sum_n |\tr U _{i}^{n}|^{2}e^{-n^2 \Delta^2 },
\end{align}
where $\rho^{fl}_{\Delta}$ is the fluctuating part of $\rho$
convoluted with a normalized gaussian of width $\Delta$. In our
computations $\Delta$ was taken large enough so that the dominant
contributions to $\rho^{fl}$ come from the $n=1$ term in the sum in
(\ref{momentsmap}).

For $\kappa=0$ the map (\ref{map}) is uniformly hyperbolic. The
perturbed map is guaranteed by Anosov's theorem to be strongly
chaotic for $\kappa \leq (\sqrt{3}-1)/\sqrt{5}\approx 0.333$. The
period-1 fixed points at
\begin{equation} \label{statpts}
q_j=\frac{1}{2}\left ( j-\frac{\kappa}{2\pi}\cos{(2\pi q_j)} \right)
\end{equation}
for integers j=0,1 are then unstable \cite{Boasman:1995}. Outside
this parameter range bifurcations can occur leading to a mixed phase
space.  When $\kappa=\kappa_{\text{bif}}=5.94338$ the phase space is
almost entirely ergodic but a saddle node bifurcation gives rise to
a new pair of period-1 orbits \cite{Berry:1998}.  For this map, it
has been shown \cite{Keating:1991, Boasman:1995} that $\tr U$ can be
expressed in the form
\begin{equation}
\tr U=\sqrt{(N/i)} \sum_{j=0}^1 \int^{\infty}_{-\infty} \exp{(2\pi i
N {\cal S}_{j}(q))}dq \label{trU}
\end{equation}
where ${\cal S}_j(q)= q^2 + \frac{\kappa}{4\pi^2} \sin{(2\pi q)}-j q$,
so that the phase is stationary at the fixed points (\ref{statpts}).

\begin{figure}
\centerline{
\includegraphics[width=8cm]{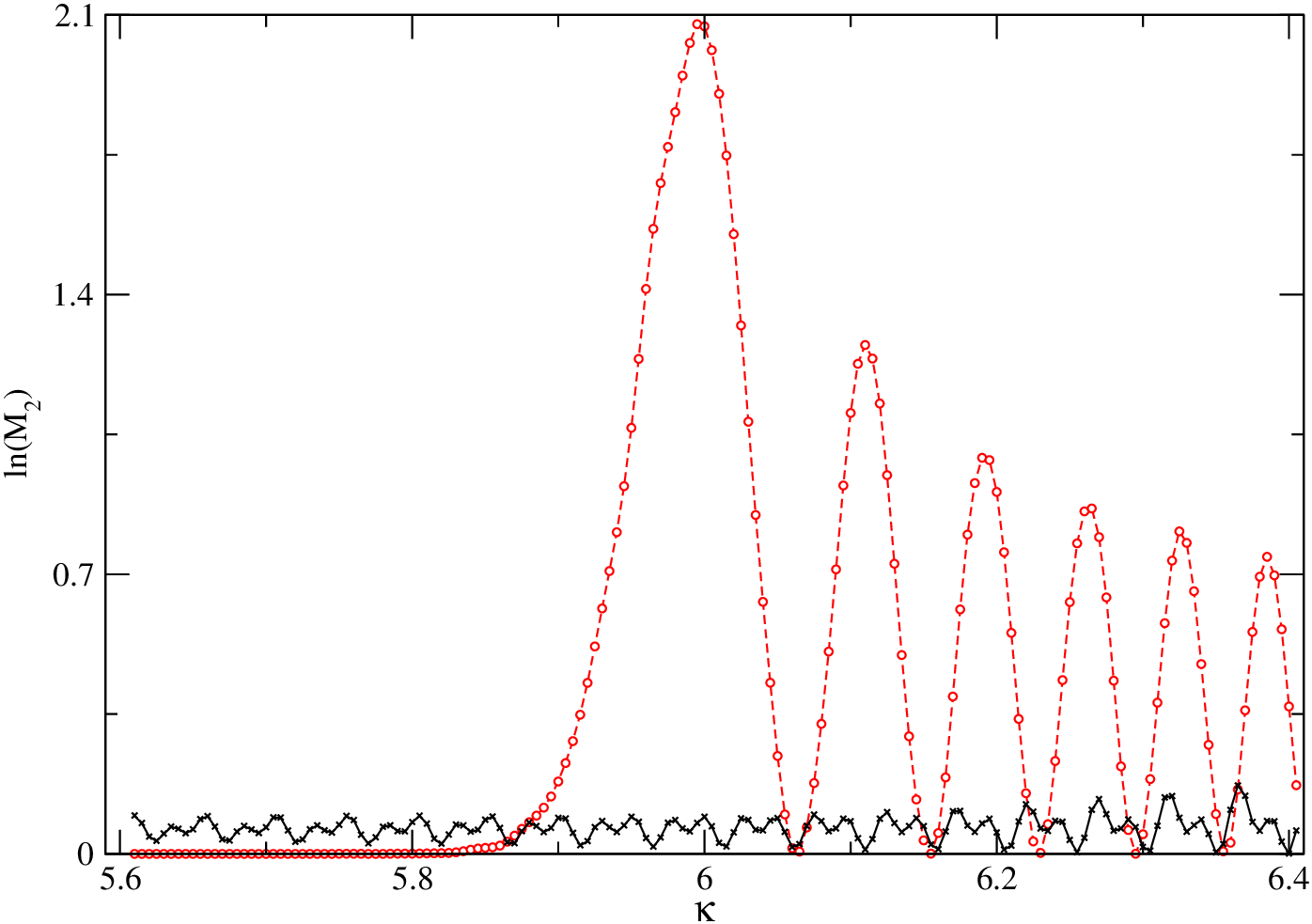}}
\caption{$\log{{\cal{M}}_2}$ versus $\kappa$ with $\Delta=1.5$ and
$N=997$. Solid black curve and crosses for strips centred at the
bifurcating fixed points; dashed red curve and red circles for
strips centred at the isolated fixed points. Curves and symbols for
theory and numerics respectively.} \label{figure1}
\end{figure}

After removing strips from phase space one gets the open map.  In
all the cases illustrated in this section, the ratio $M/N$ of the
dimension of the quantum map to the number of open channels is 0.28.
In order to see the contribution of bifurcating points to the
moments (\ref{momentsmap}) we have computed $|\tr U_{II}|^2$ in two
different configurations.  First, two strips are located in such a
way that they block the unstable fixed points $q_0=0.81425$ and
$q_1=0.6857$. Second, the two strips are placed such that they block
the bifurcating points $q_0^+=0.4453$ and $q_1^+=0.0548$. These are
period-1 fixed points which originate in a saddle node bifurcation
at $\kappa_{\text{bif}}$.  The logarithm of the second moment
${\cal{M}}_2$ is plotted in figure \ref{figure1} as a function of
the perturbation parameter $\kappa$. The two strips block either the
bifurcating points (solid black curve and crosses) or the isolated
points (dashed red curve and circles).  In this and the subsequent
two figures the theoretical curves correspond to asymptotic
evaluations of (\ref{trU}) \cite{Berry:1998}, as described in
outline in the previous section.

\begin{figure}
\centerline{
\includegraphics[width=8cm]{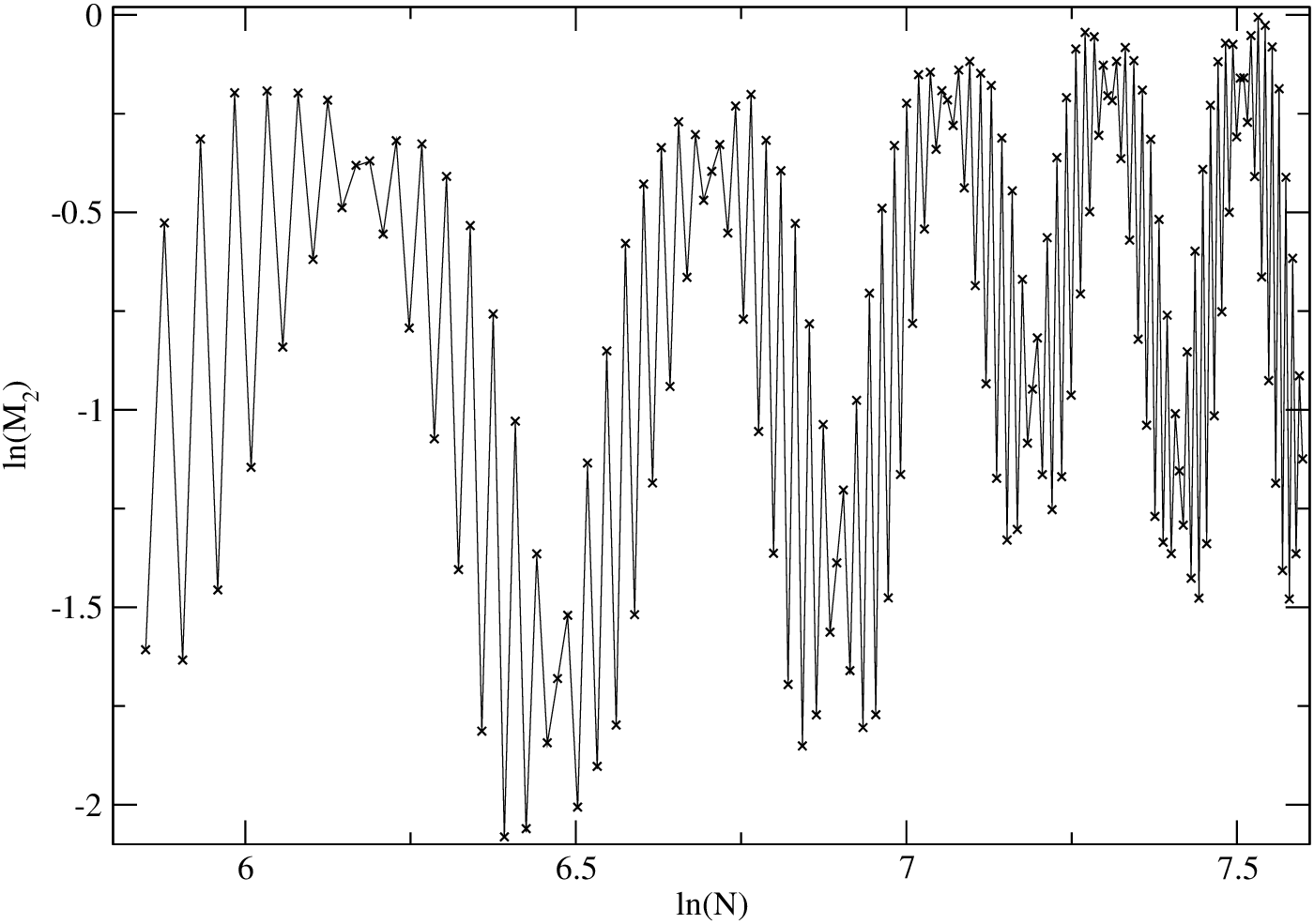}}
\caption{$\log{{\cal{M}}_2}$ versus $\log{N}$ for $\Delta=1.5$ and
$\kappa=\kappa_{\text{bif}}$.  Strips centred at the bifurcating
fixed points. Solid black curve and crosses for theory and direct
numerical evaluation of the trace using (\ref{momentsmap})
respectively.} \label{figure2}
\end{figure}

Figure \ref{figure2} illustrates the contribution of isolated points
to ${\cal{M}}_2$ as a function of the dimension $N$. In this case,
strips are centred at the bifurcating points so that only the
isolated fixed points contribute and hence $|\tr U|^2$ is of order 1
\cite{Berry:1998}.

\begin{figure}
\centerline{
\includegraphics[width=8cm]{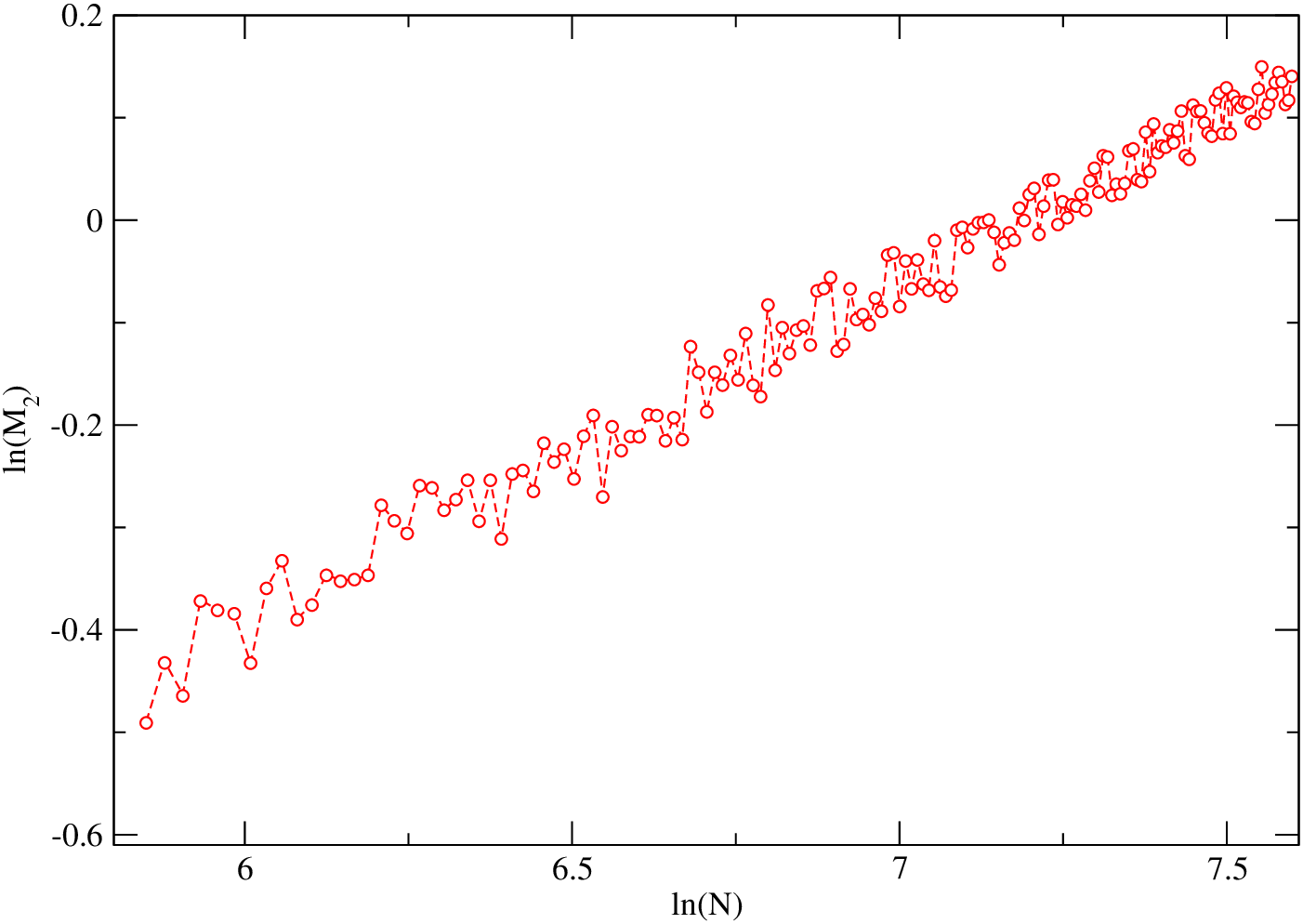}}
\caption{$\log{\cal{M}}_2$ versus $\log{N}$ for $\Delta=1.5$ and
$\kappa=\kappa_{\text{bif}}$. Strips centred at the isolated fixed
points. Dashed red curve and circles for theory and direct numerical
evaluation of the trace using (\ref{momentsmap}) respectively.}
\label{figure3}
\end{figure}

By contrast, figure \ref{figure3} illustrates the situation where
the strips block the isolated orbits. The contribution of the
remaining, bifurcating orbits to $|\tr U|^2$ grows semiclassically
like $N^{1/3}$ in this case \cite{Berry:1998} (i.e.~the scaling
exponent is 1/3).  For $\kappa$ around $\kappa_{\text{bif}}$, as in
figure \ref{figure1}, $\tr U_{II}$ is more intricate and involves
Airy functions \cite{Berry:1998}.

These figures illustrate clearly how the bifurcating trapped
periodic orbit dominates the fluctuations of the time delay
semiclassically.

\section*{Acknowledgements}

We are grateful to the Royal Society for funding this collaboration.

\end{document}